\documentclass[twocolumn,superscriptaddress,aps,preprintnumbers,amsmath,amssymb,prl,nofootinbib]{revtex4-2}
\pdfoutput=1

\usepackage{graphicx,rotating} 
\usepackage[bookmarksopen,colorlinks=true,linkcolor=steelblue,citecolor=darkred,urlcolor=darkred,linktoc=all]{hyperref}

\usepackage{mathtools}
\usepackage{slashed}
\usepackage{bm}
\usepackage{bbm}
\usepackage{comment}
\usepackage[utf8]{inputenc}
\usepackage{accents}
\usepackage{cancel}
\usepackage{physics}
\usepackage{here}

\usepackage{libertine}
\usepackage[ISO, upint]{libertinust1math}
\usepackage[scr=boondoxo,bb=boondox]{mathalpha}
\usepackage[T1]{fontenc}

\usepackage{stmaryrd}
\usepackage{trimclip}
\usepackage[framemethod=default]{mdframed}
\newmdenv[skipabove=7pt,
skipbelow=7pt,
rightline=false,
leftline=false,
topline=false,
bottomline=false,
backgroundcolor=gray!10,
linecolor=gray,
innerleftmargin=5pt,
innerrightmargin=5pt,
innertopmargin=5pt,
innerbottommargin=5pt,
leftmargin=0cm,
rightmargin=0cm,
linewidth=4pt]{eBox}
\newmdenv[skipabove=7pt,
skipbelow=7pt,
rightline=true,
leftline=true,
topline=true,
bottomline=true,
backgroundcolor=white,
linecolor=gray,
innerleftmargin=5pt,
innerrightmargin=5pt,
innertopmargin=5pt,
innerbottommargin=5pt,
leftmargin=0cm,
rightmargin=0cm,
linewidth=1pt]{eBox2}

\usepackage{xcolor}
\definecolor{darkred}{rgb}{0.7, 0., 0.}
\definecolor{steelblue}{rgb}{0.275,0.51, 0.706}

\begin{document}

\title{Revisiting constraints on magnetogenesis from baryon asymmetry}

\author{Yuta Hamada}
\email{yhamada@post.kek.jp}
\affiliation{Theory Center, IPNS, KEK, 1-1 Oho, Tsukuba, Ibaraki 305-0801, Japan}
\affiliation{Graduate University for Advanced Studies (Sokendai), 1-1 Oho, Tsukuba, Ibaraki 305-0801, Japan}

\author{Kyohei Mukaida}
\email{kyohei.mukaida@kek.jp}
\affiliation{Theory Center, IPNS, KEK, 1-1 Oho, Tsukuba, Ibaraki 305-0801, Japan}
\affiliation{Graduate University for Advanced Studies (Sokendai), 1-1 Oho, Tsukuba, Ibaraki 305-0801, Japan}

\author{Fumio Uchida}
\email{fumiouchida@ibs.re.kr}
\affiliation{Theory Center, IPNS, KEK, 1-1 Oho, Tsukuba, Ibaraki 305-0801, Japan}
\affiliation{Kavli IPMU (WPI), UTIAS, University of Tokyo, 5-1-5 Kashiwanoha, Kashiwa, Chiba 277-8583, Japan}
\affiliation{CTPU-CGA, Institute for Basic Science (IBS), Daejeon, 34126, Korea}

\date{\today}

\begin{abstract}
Magnetic fields in the universe potentially serve as a messenger of primordial physics.
The observationally suggested intergalactic magnetic fields may be a relic of helical primordial $\mathrm{U}(1)_Y$ magnetic fields, which may also explain the origin of the baryon asymmetry of the universe.
This scenario has been considered to be not viable, which we revisit as well as the baryon isocurvature problem for non-helical primordial $\mathrm{U}(1)_Y$ magnetic fields, based on the recent discussion on the hot electroweak theory.
We find that maximally helical fields can be the origin of both the intergalactic magnetic fields and the baryon asymmetry of the universe and that there can be a window for non-helical fields to explain the origin of the intergalactic magnetic fields if the Higgs dynamics during the electroweak crossover compensate the helicity decay with a $\lesssim10^{-9\text{--}-10}$ precision.
\end{abstract}

\maketitle

\noindent\textit{\textbf{Introduction.---\,}}
Magnetic fields are observed across a wide range of scales from compact stars to galaxies, galaxy clusters, and even the cosmic voids.
However, their origin remains an open question.
An intriguing possibility is that they have a common origin in the early universe, potentially serving as a messenger of primordial physics.
If this is the case, a particularly interesting signal is the presence of intergalactic magnetic fields (IGMFs).
The primordial magnetic fields at the recombination epoch may be unaffected by the structure formation and astrophysical processes and survive in the cosmic voids today~\cite{Durrer:2013pga}. 

Observationally, gamma-ray spectra of $\mathrm{TeV}$ blazars encode to what extent the IGMFs let the high-energy electron-positron pairs in the cascade process from $\mathrm{TeV}$ down to $\mathrm{GeV}$ away from the line of sight \cite{Honda1989,Plaga:1995ins}.
The recent observations show the lack of secondary GeV photons, which may suggest the presence of IGMFs~\cite{Neronov:2010gir,Tavecchio:2010mk, Tavecchio:2010ja,Dolag:2010ni,Essey:2010nd,Taylor:2011bn,Dermer:2010mm,Vovk:2011aa,Takahashi:2013lba,Finke:2015ona,Fermi-LAT:2018jdy,AlvesBatista:2020oio,MAGIC:2022piy} (see Refs.~\cite{Broderick:2011av,Schlickeiser:2012vog,Schlickeiser:2013eca,Chang:2014cta,Vafin:2018kox} for counterarguments).
Assuming that this depletion is indeed due to IGMFs,
they have obtained more or less similar estimation for the amplitude of magnetic fields $B_\mathrm{void}$ and its correlation length in the cosmic voids $\xi_{\mathrm{M,void}}$:
\begin{equation}
    B_\mathrm{void}>10^{-17}\,\mathrm{G}\cdot\mathrm{max}\bigg\{\qty(\dfrac{\xi_{\mathrm{M,void}}}{0.2\,\mathrm{Mpc}})^{-\frac{1}{2}},1\bigg\}.\label{eq:blazar}
\end{equation}

The proposed scenarios of primordial magnetogenesis~\cite{Turner:1987bw,Ratra:1991bn,Garretson:1992vt,Hogan:1983zz,Quashnock:1988vs,Cheng:1994yr,Vachaspati:1991nm} are classified into two categories based on the time of their operation, namely before or after the electroweak symmetry breaking (EWSB).
The main focus of this work is magnetogenesis operating before EWSB.
In this case, the $\mathrm{U}(1)_Y$ magnetic fields at the cosmological scale are generated by a certain mechanism, \footnote{
    The $\mathrm U(1)_Y$ magnetic fields cannot be cut thanks to the magnetic one-form symmetry.
    Since the $\mathrm{SU} (2)_\mathrm{L}$ magnetic fields can be cut because of the lack of the magnetic one-form symmetry, their production are not expected to contribute to the cosmological relics.
} 
which later turns into the $\mathrm{U}(1)_\mathrm{em}$ ones during the electroweak crossover.
The crossover nature of EWSB implies that the magnetic one-form symmetry is always in the spontaneously broken phase.
The associated Nambu--Goldstone boson is the ``magnetic'' field, which smoothly changes from the $\mathrm{U}(1)_Y$ one before EWSB to the $\mathrm{U}(1)_\mathrm{em}$ one after EWSB.
In this way, the primordial $\mathrm{U}(1)_Y$ magnetic fields are converted to the $\mathrm{U}(1)_\mathrm{em}$ ones.

In the literatures, it has been discussed that this conversion process can have a direct relation to the baryon asymmetry~\cite{Kamada:2016cnb,Kamada:2020bmb}.
Indeed, the Sakharov conditions~\cite{Sakharov:1967dj} are fulfilled since the presence of cosmological magnetic fields break both $C$ and $CP$~\cite{Davidson:1996rw} and the conversion is out-of-equilibrium process.
Through the Adler--Bell--Jackiw (ABJ) anomaly in the Standard Model (SM) \cite{Adler:1969gk,Bell:1969ts,tHooft:1974kcl}, the baryon asymmetry is generated if the $\mathrm{U}(1)_Y$ magnetic helicity or the $\mathrm{SU}(2)_\mathrm{L}$ Chern--Simons number change during the conversion process.
Since the baryon asymmetry of the universe (BAU) is observationally determined \cite{Planck:2018vyg} and  the primordial deuterium abundance constrains baryon isocurvature perturbations \cite{Inomata:2018htm}, the primordial $\mathrm{U}(1)_Y$ magnetic field can be constrained in light of the baryon asymmetry \cite{Joyce:1997uy,Giovannini:1997eg,Giovannini:1997gp,Fujita:2016igl,Kamada:2016eeb,Kamada:2016cnb,Kamada:2020bmb,Boyer:2025jno}.

In particular, Ref.~\cite{Kamada:2020bmb} concluded that it is unlikely for the primordial $\mathrm{U}(1)_Y$ magnetic field to explain the observed strength of the IGMFs, Eq.~\eqref{eq:blazar}, without spoiling the standard Big-Bang Nucleosynthesis (BBN).

Recently, the authors of this work have discussed the fate of the magnetic field during the electroweak crossover and pointed out crucial updates to the theoretical understanding~\cite{Hamada:2025cwu}.
The interplay between the magnetic field and the baryon asymmetry has turned out to involve a qualitative uncertainty, which was not recognized in the previous literatures.
A particularly significant one is the role of the Higgs field.
It is in principle possible that the conversion process by itself would not change the $\mathrm{U}(1)_Y$ magnetic helicity and the $\mathrm{SU}(2)_\mathrm{L}$ Chern--Simons number at all because of the Higgs-field dynamics, which significantly suppresses the baryon asymmetry generation.

In this work, we discuss how the refined understanding in Ref.~\cite{Hamada:2025cwu} affect the baryon asymmetry constraints on the primordial $\mathrm{U}(1)_Y$ magnetic field and whether they can be the origin of the observed IGMFs while fulfilling the updated constraints.
Our discussion indicates that the conclusion of Ref.~\cite{Kamada:2020bmb} may not hold once previously unrecognized theoretical uncertainties are taken into account.
In particular, we discuss an intriguing possibility that the IGMFs and the BAU share the same origin, namely the helical $\mathrm{U}(1)_Y$ primordial magnetic field.

\noindent\textit{\textbf{Baryogenesis from helicity decay.---\,}}
We first review the recent discussion on how the magnetic helicity decay during the electroweak crossover can generate the baryon number.

Suppose that there exist $\mathrm{U}(1)_Y$ primordial magnetic fields on cosmological length scales at high temperatures well above the electroweak scale, $T\gg T_\mathrm{EW}=\mathcal O(100)\,\mathrm{GeV}$.
We do not specify the mechanism of magnetogenesis, and we will keep the discussion general.
The $\mathrm{U}(1)_Y$ magnetic fields can contribute to the macroscopic physics is because they cannot be screened by local processes.
Namely, the $\mathrm{U}(1)_Y$ (anti)monopoles are absent in the thermal plasma after inflation, since otherwise we would suffer from the notorious monopole problem.
This means that the Bianchi identity, $\vec\nabla\cdot\vec B_Y=0$, holds, implying that the SM enjoys the magnetic one-form symmetry $\mathrm{U}(1)_\mathrm{M}^{[1]}$ associated with the $\mathrm{U}(1)_Y$ gauge field.
It is known that the magnetic symmetry is spontaneously broken in the high-temperature plasma phase, whose Nambu--Goldstone boson is the $\mathrm U(1)_Y$ magnetic field.

As the temperature of the universe drops much below the electroweak scales, the macroscopic magnetic field cannot continue being the $\mathrm{U}(1)_Y$ one because the Higgs condensate $\Phi^\dag \Phi$ obtains its vacuum expectation value (VEV), and thereby the SM gauge fields, including the $\mathrm U(1)_Y$ gauge field, involves not only the massless modes but the massive modes.
At sufficiently low temperatures, the massless (unconfined) mode is identified as the $\mathrm{U}(1)_\mathrm{em}$ one, namely a particular linear combination of $\vec B_\mathrm{em}=\cos\theta_\mathrm{w}\vec B_Y-\sin\theta_\mathrm{w}\vec B_{W^3}$ with $\theta_\mathrm{w}$ being the weak mixing angle.
On the other hand, the $Z$-magnetic flux, $\vec B_Z=\sin\theta_\mathrm{w}\vec B_Y+\cos\theta_\mathrm{w}\vec B_{W^3}$, is massive and hence stabilized to zero.

The transition between these two phases have been a subject of lattice studies, which indicate a crossover transition \cite{Kajantie:1996mn,Gurtler:1997hr,Rummukainen:1998as,Csikor:1998eu,Aoki:1999fi,DOnofrio:2015gop}.
From the symmetry viewpoint, this suggests that the order parameters of the SM global symmetries remain in the same phase at all temperatures across the electroweak crossover.
Indeed, the breaking patterns of generalized global symmetries in the three-dimensional hot electroweak theory are the same at the high- and low- temperatures \cite{Hamada:2025cwu}.
In particular, the spontaneously broken magnetic symmetry\footnote{
    Precisely speaking, the one-form magnetic symmetry in the original $1$+$3$d theory is split into $\mathrm{U}(1)_\mathrm{M}^{[0]}$ and $\mathrm{U}(1)_\mathrm{M}^{[1]}$ symmetries in the three-dimensional effective field theory.
    The former exhibits a spontaneous symmetry breaking (SSB) while the latter remains unbroken at all temperatures.
} implies a massless Nambu--Goldstone boson, which corresponds to the unconfined (denoted as $\mathrm{U}(1)_{\rm c\mkern-7.5mu/}$) magnetic field.
Provided that the magnetic symmetry remains in the SSB phase at all temperatures, the Nambu--Goldstone mode interpolates the $\mathrm{U}(1)_Y$ and the $\mathrm{U}(1)_\mathrm{em}$ magnetic fields at high- and low- temperatures.

The Nambu--Goldstone mode may be imprinted in a certain linear combination of gauge invariant operators, \textit{i.e.,} $\vec{B}_Y$ and $\Phi^\dag \vec{B}_W \Phi$.
It is convenient to define a gauge-independent effective mixing angle $\theta_\mathrm{eff}$ of these operators so that its orthogonal direction is completely gapped~\cite{Hamada:2025cwu}
\begin{equation}
    \cos\theta_{\rm eff}(T)
        \coloneq\qty(\!1+\dfrac{g_3^2(T)\sin^2\theta_{\rm w}}{4\pi m_W(T)}\!)\cos\theta_{\rm w},
    \label{eq:coseff}
\end{equation}
at the one-loop level for a relatively low temperatures, $100\,\mathrm{GeV}<T<150\,\mathrm{GeV}$.
Here the three-dimensional gauge coupling is related to the original one via $g_3^2\coloneq g^2 (T) T$, and $m_W(T)=g_3(T)v_3(T)/2$ is the $W$ boson mass determined by the Higgs VEV $v_3(T)$.\footnote{Our $\theta_{\rm eff}$ is different from the ``effective'' weak mixing angle $\theta_{\rm eff}^\mathrm{KL}$ introduced earlier in Refs.~\cite{DOnofrio:2015gop,Kamada:2016cnb}:
\begin{align}
    \cos^2\!\theta_{\rm eff}^\mathrm{KL}\,(T)
        =\qty(\!1+\dfrac{11g_3^2(T)\sin^2\theta_{\rm w}}{12\pi m_W(T)}\!)\cos^2\!\theta_{\rm w}.
    \label{eq:coseff_KL}
\end{align}
Their definition does not take into account the wavefunction renormalization factor for the unconfined magnetic field, and hence we encounter an inconsistency when we compare the massless-pole residues of the $Y$--$Y$ and the $W^3$--$W^3$ correlators.
Moreover, if we try to disentangle contributions from the renormalization factor and the effective weak mixing angle in Eq.~\eqref{eq:coseff_KL}, the result depends on the gauge choice \cite{Hamada:2025cwu}.
}

The time-dependent effective mixing angle indicates how the primordial ${\mathrm U}(1)_Y$ magnetic fields gradually transition to the $\mathrm{U}(1)_\mathrm{em}$ ones.
In the literatures~\cite{Joyce:1997uy,Giovannini:1997eg,Giovannini:1997gp,Fujita:2016igl,Kamada:2016eeb,Kamada:2016cnb,Kamada:2020bmb}, it was conjectured that this conversion process is accompanied by the generation of the baryon asymmetry through the ABJ anomaly in the SM \cite{Adler:1969gk,Bell:1969ts,tHooft:1974kcl}
\begin{equation}
    \Delta Q_{B+L} = 3 \cdot 2\,  \qty(\Delta N_{\rm CS}^{\mathrm{SU}(2)_{\mathrm L}}  - \Delta H_Y ),
\end{equation}
where $Q_{B+L}$ is the comoving $B+L$ charge, $N_\mathrm{CS}^{\mathrm{SU}(2)_\mathrm{L}}$ is the $\mathrm{SU}(2)_\mathrm{L}$ Chern--Simons number, and $H_Y$ is the $\mathrm{U}(1)_Y$ magnetic helicity.
To see how the conversion process can generate the baryon asymmetry, let us recall that the $\mathrm U(1)_Y$ magnetic flux is never screened.
Hence any changes of $H_Y$ must be accompanied by macroscopic processes such as reconnection or contraction of the $\mathrm U(1)_Y$ magnetic fluxes, which is however suppressed by the large electric conductivity.
In contrast, the $\mathrm{SU}(2)_{\mathrm L}$ magnetic fluxes can vary by local processes.

Motivated by this observation, one possible dynamics across the crossover would be the following.
After a short time interval $\Delta t$, the originally unconfined magnetic field turns into a superposition of the confined (denoted by a subscript ``$\mathrm{c}$'') and the unconfined (``${\rm c\mkern-7.5mu/}$'') ones, and then the confined magnetic fields become the unconfined ones by dressing the $\mathrm{SU}(2)_\mathrm{L}$ gauge fields. 
Through this process, the $\mathrm{SU}(2)_\mathrm{L}$ gauge fields can acquire a non-zero Chern--Simons number, and hence the $B+L$ asymmetry is generated~\cite{Hamada:2025cwu}
\begin{align}
    \label{eq:B+L}
    \Delta Q_{B+L}&\simeq
    3 \cdot 2\,  \cot^2 \theta_\text{w}\, \Delta\qty[ \qty( \tan^2 \theta_\text{eff} - \tan^2 \theta_\text{w} )  H_{\rm c\mkern-7.5mu/}]\notag\\
    &\quad+ 3 \cdot 2\, \Delta N_{\rm CS}^{\mathrm{SU}(2)_{\mathrm L}} \Big|_\text{NP},
\end{align}
where $H_{\rm c\mkern-7.5mu/} \coloneq (g'^2 \cos^2 \theta_{\rm eff}/16 \pi^2) \int \dd^3 x\, \vec{A}_{\rm c\mkern-7.5mu/} \cdot \vec{B}_{\rm c\mkern-7.5mu/}$ is the comoving magnetic helicity of the unconfined magnetic flux.
While the unconfined helicity, $H_{\rm c\mkern-7.5mu/}$, is constant up to dissipation effects, reflecting the conservation of $H_Y$, the confined helicity decay is imprinted in $\Delta\theta_{\mathrm{eff}}$.
The non-perturbative contribution is denoted as $\Delta N_{\rm CS}^{\mathrm{SU}(2)_{\mathrm L}} \big|_\text{NP}$.
If the electroweak sphaleron is the only non-perturbative contribution, an approximate equilibrium between the helicity decay and the sphaleron $B+L$ washout establishes, and the equilibrium solution at the sphaleron freezeout $T_\mathrm{fo}\sim 130\,\mathrm{GeV}$ \cite{DOnofrio:2014rug} gives a rough estimate of the resultant $B+L\,(\sim 2B\sim 2L)$ abundance \cite{Kamada:2016cnb}.

However, this is not the only process allowed by the system. Without generating the $\mathrm{SU}(2)_\mathrm{L}$ gauge fields, the Higgs dynamics can convert the confined magnetic flux into the unconfined flux in accordance to the $\mathrm{U}(1)^{[1]}_\mathrm{M}$ symmetry.
An illustrative example is untying a link of the confined magnetic fluxes by producing the Nambu monopole pairs on the flux, which cut the confined flux. 
Without intersecting magnetic fluxes, the confined field can never develop finite $\vec E_\mathrm{c}\cdot\vec B_\mathrm{c}$ which could contribute to $\partial_\mu j_{B+L}^\mu$ through the anomaly equation, implying that the confined helicity decay does not necessarily generate baryon number.
If this is the case, the resultant $B+L$ asymmetry is solely sourced by the unconfined helicity decay 
through the dissipation~\cite{Hamada:2025cwu,Fukuda:2025nmc}:
\begin{equation}
    \label{eq:B+L_np}
    \hspace{-2.5mm}
    \Delta^\text{diss} Q_{B+L} \simeq 
    3 \cdot 2\,\qty(\dfrac{\tan^2 \theta_\text{eff}}{\tan^2 \theta_\text{w}}-1) \Delta H_{\rm c\mkern-7.5mu/} + 3 \cdot 2\, \Delta N_{\rm CS}^{\mathrm{SU}(2)_{\mathrm L}} \Big|_\text{sph}.
\end{equation}

These two estimations can be regarded as the two extreme relaxation processes in terms of the gauge invariant indicator $\delta = H_Y - N_{\rm CS}^{\mathrm{SU}(2)_{\mathrm L}} + N_H$ with $N_H$ being the Higgs winding number.
Initially, we have non-vanishing $\delta$ due to the non-zero $H_Y$, while the groundstate in the presence of unconfined magnetic flux is characterized by $\delta = 0$ \cite{Fukuda:2025nmc}.
The first estimation \eqref{eq:B+L} with the electroweak sphaleron process as the only NP contribution corresponds to the relaxation by changing $N_{\rm CS}^{\mathrm{SU}(2)_{\mathrm L}}$, while the second one \eqref{eq:B+L_np} changes $N_H$.

To draw a definite conclusion, we need to perform a numerical simulation, which is beyond the scope of this work.
In the following, we will discuss how this uncertainty affects the constraints on the primordial magnetic field from the baryon asymmetry.

\noindent\textit{\textbf{Net baryon asymmetry generation.---\,}}%
Based on the preceding discussion, we will give a formula that relates the primordial magnetic field and the baryon asymmetry.
We here limit ourselves to discuss exclusively the homogeneous component of a given local variable $X$, namely its average over an infinitely-large volume, $\overline{X}\coloneq \lim_{V\to\infty}V^{-1}\int_V\dd^3xX$,
where one may safely define the averaged magnetic helicity in a gauge invariant manner owing to the lack of magnetic monopoles. 
We also assume that the primordial $\mathrm U(1)_Y$ magnetic field carries the net magnetic helicity.

We begin with the unconfined helicity dissipation.
As can be seen from Eq.~\eqref{eq:B+L_np}, the decay of $H_{\rm c\mkern-7.5mu/}$ sources $B+L$ even if the Higgs dynamics completely compensates the confined helicity decay. 
In analogy with the ordinary magneto-hydrodynamics in an expanding spacetime \cite{Brandenburg:1996fc}, we assume that the unconfined magnetic flux is dissipated proportionally to the comoving resistivity $\sigma_{\rm c\mkern-7.5mu/}^{-1}$.
By balancing this source with the sphaleron washout,\footnote{Just before the electroweak sphaleron freezes out, namely at $145\,\mathrm{GeV}>T>T_\mathrm{fo}$, $\Gamma_\mathrm{sph}$ determines the rate of $B+L$ violation because the other relevant interactions such as the electron spin-flip are faster \cite{Kamada:2016cnb}.} $-\Gamma_\mathrm{sph}Q_{B+L}\Delta t$, where $\Gamma_\mathrm{sph}$ is the sphaleron rate, we obtain
\begin{align}
    \!\!\!\!\!\!\overline{n_{B+L,\mathrm{fo}}^\mathrm{diss}}
        =
    &
        \frac{3 \cdot 2}{16 \pi^2a_\mathrm{fo}(\dot a/a)_\mathrm{fo}}\Big[\Big(\dfrac{\tan^2 \theta_\text{eff}}{\tan^2 \theta_\text{w}}-1\Big)
    \notag\\
    &
        \hspace{2mm}\cdot g'^2\cos^2\theta_\mathrm{eff}\qty(-2\sigma_{\rm c\mkern-7.5mu/}^{-1}\overline{\vec{B}_{\rm c\mkern-7.5mu/}\cdot\vec\nabla\times\vec{B}_{\rm c\mkern-7.5mu/}})\,\Big]_\mathrm{fo\!}.\!\!\!\!
        \label{eq:nB+L_3}
\end{align}
In this expression, comoving density of $B+L$ charge, $n_{B+L}$, is evaluated at the sphaleron freezeout, $\Gamma_\mathrm{sph}(T_\mathrm{fo})\simeq (\dot a/a)_\mathrm{fo}$, 
where $a$ is the scale factor of the universe, and $(\dot a/a)_\mathrm{fo}$ is the Hubble constant at the freezeout temperature.

In addition to this, there exists the contribution from the confined helicity decay if the decay is not dominated by the Higgs dynamics.
Similary, by balancing this source and the sphaleron washout, we obtain
\begin{align}
    \overline{n_{B+L,\mathrm{fo}}^\mathrm{cdec}}
        =\alpha\frac{3 \cdot 4\,g'^2 \cot^2 \theta_{\rm w}}{16 \pi^2 (\dot a/a)_\mathrm{fo}}
        \Big[\dot\theta_\text{eff} \, \tan \theta_\text{eff} \,\overline{\vec{A}_{\rm c\mkern-7.5mu/} \cdot \vec{B}_{\rm c\mkern-7.5mu/}}\,\Big]_\mathrm{fo},
        \label{eq:nB+L_1}
\end{align}
where we include an efficiency factor $0\leq\alpha\leq1$ to account for the possible Higgs dynamics that could compensate the confined helicity decay.
The confined helicity decay only via the change of $N_H$ corresponds to $\alpha = 0$ while that only via the change of $N_{\rm CS}^{\mathrm{SU}(2)_{\mathrm L}}$ does to $\alpha = 1$.

In total, the primodial helical magnetic field results in a net baryon asymmetry, $\overline{n_{B+L,\mathrm{fo}}^\mathrm{diss}}+\overline{n_{B+L,\mathrm{fo}}^\mathrm{cdec}}$.
To evaluate this, we parametrize the degree of parity-violation of the unconfined magnetic field with a single parameter $\epsilon,\;\vert\epsilon\vert\leq1$.
Consequently, we obtain 
\begin{align}
    \overline{\eta_{B,\mathrm{fo}}}
        =&\,5\cdot10^{-11}\,\alpha\epsilon_\mathrm{i}\qty(\dfrac{B_\mathrm{i}}{10^{-17}\,\mathrm{G}})^{\!\!2}\qty(\dfrac{\xi_\mathrm{M,\,i}}{10^{-9}\,\mathrm{Mpc}})
    \notag\\[5pt]
        &
        \,+8\cdot10^{-10}\,\epsilon_\mathrm{i}\qty(\dfrac{B_\mathrm{i}}{10^{-7}\,\mathrm{G}})^{\!\!2}\qty(\dfrac{\xi_\mathrm{M,\,i}}{10^{-9}\,\mathrm{Mpc}})^{\!\!-1},\!\!\!
    \label{eq:diss_net}
\end{align}
where the subscript $_\mathrm{i}$ indicates the initial condition at the electroweak scale for the subsequent MHD evolution, which is non-trivial even in terms of comoving quantities (See Appendix).
$B_\mathrm{i}$ is the comoving magnetic field strength, $\xi_\mathrm{M,\,i}$ is the comoving magnetic coherence length, and
$\eta_B(\vec x,T)\coloneq n_B(\vec x,T)/s(T)$ is the baryon-to-entropy ratio.
In this expression, $n_B=(1/2)n_{B+L}$ is the comoving density of baryon number, and $s$ is the comoving entropy density.
The first (second, {\it resp.}) line is the contribution from confined helicity decay (unconfined helicity dissipation, {\it resp.}).

\begin{figure}[t]
    \includegraphics[keepaspectratio, width=.44\textwidth]{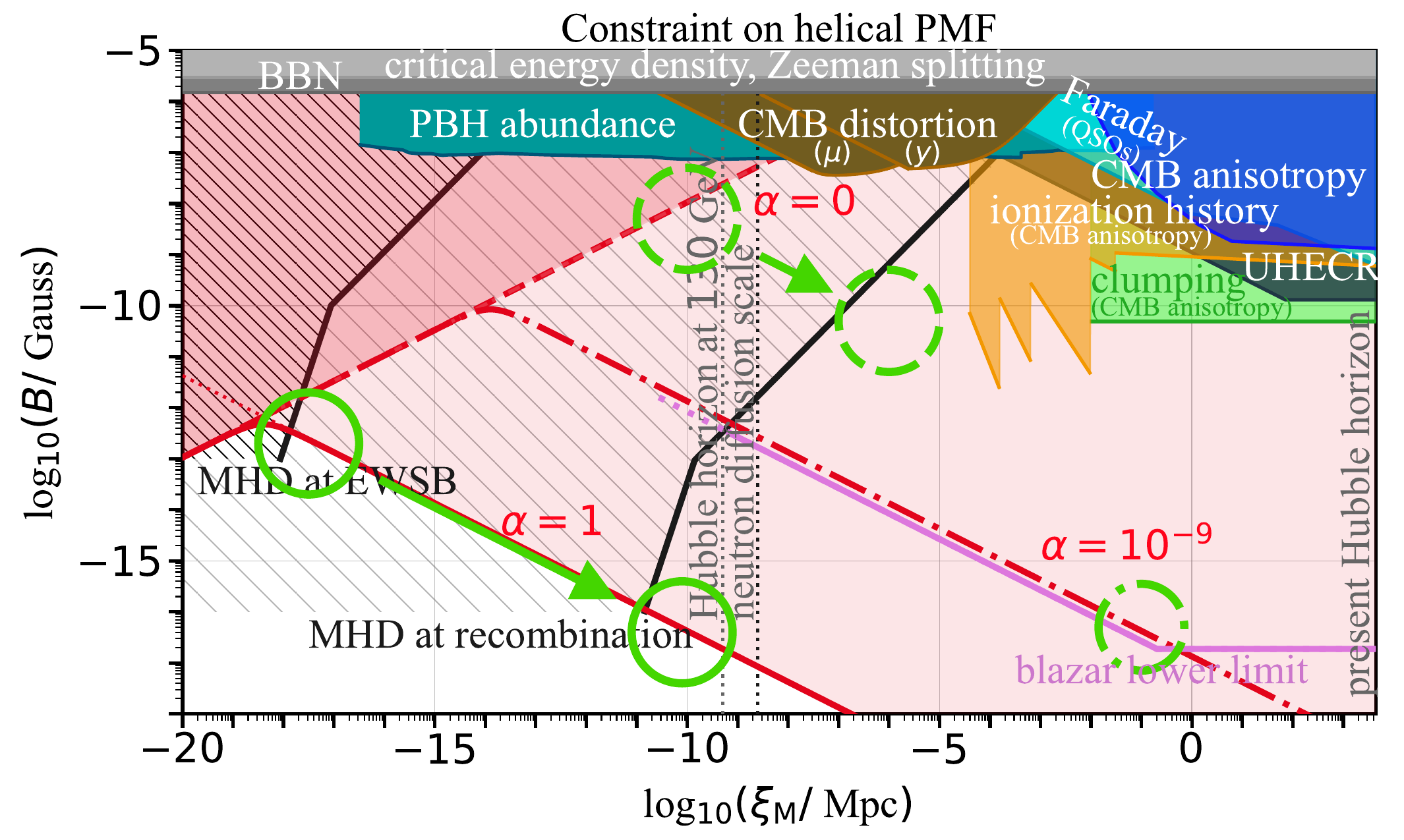}
    \includegraphics[keepaspectratio, width=.44\textwidth]{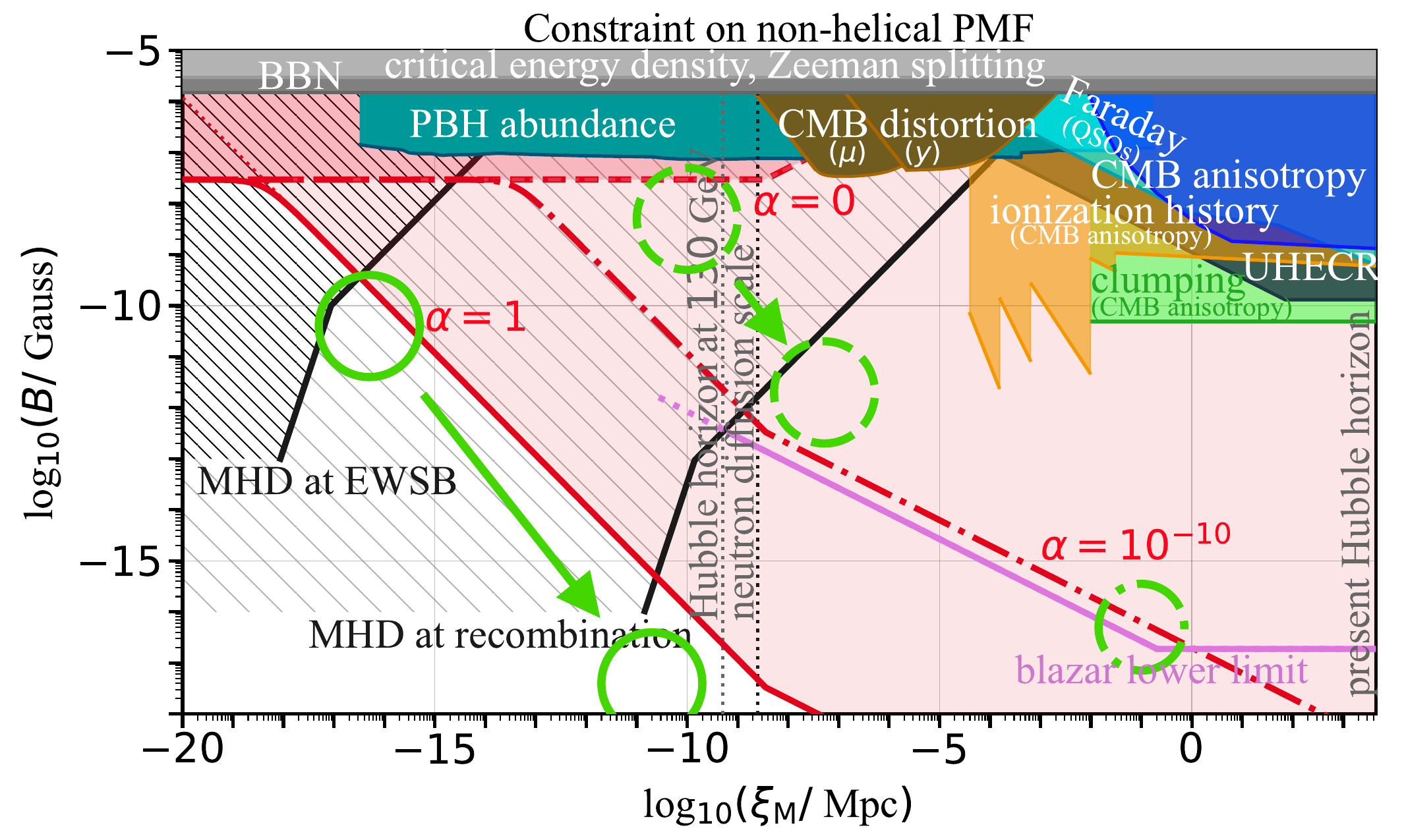}
    \caption{
    {\it Top}: the baryon-overproduction constraint for maximally helical ($\epsilon_\mathrm{i}=1$). {\it Bottom}: the baryon-isocurvature constraint for non-helical ($\vert\epsilon_\mathrm{i}\vert\ll1$) primordial magnetic field at $T_\mathrm{fo}\simeq130\,\mathrm{GeV}$ (red-shaded regions) and the blazar lower bound \eqref{eq:blazar} on the void magnetic field in the present universe (purple-solid line).
    The red-solid lines represent the baryon asymmetry constraints for $\alpha=1$, the red-dashed ones for $\alpha=0$, and the red dash-dot ones for $\alpha=10^{-9}/10^{-10}$.
    The green circles show three possible examples of primordial magnetic fields (see the main text).
    For reference, the black-hatched regions are excluded by magneto-hydrodynamics \cite{Uchida:2024ude}, and the color-shaded regions are several known upper bounds \cite{Kawasaki:2012va,Kushwaha:2024zhd,Uchida:2024iog,Planck:2015zrl,Paoletti:2022gsn,Jedamzik:2018itu,Blasi:1999hu,Neronov:2021xua,Neronov:2009gh} (See Appendix A of Ref.~\cite{Uchida:2024ude} for details). Zoom-up figures are shown in Appendix.
    }
    \label{fig:blazar}
\end{figure}

Now we come to the baryon-overproduction constraint on the primordial magnetic field \cite{Fujita:2016igl,Kamada:2016cnb,Kamada:2016eeb}.
Assuming we have no other (anti-)baryon generation mechanisms, we impose
\begin{align}
    \overline{\eta_{B,\mathrm{fo}}}\leq\overline{\eta_B^\mathrm{obs}}
    \label{eq:BO}
\end{align}
where $\overline{\eta_B^\mathrm{obs}}=9\times10^{-11}$ is the observed baryon asymmetry of the universe \cite{Planck:2018vyg}.
The constraint for the maximally helical case, $\epsilon_\mathrm{i}=1$, is shown in {\it Top} panel of Fig.~\ref{fig:blazar}.
In particular, when the equality in Eq.~\eqref{eq:BO} holds, the helical primordial magnetic field can be the origin of the BAU.

\noindent\textit{\textbf{Local baryon asymmetry generation.---\,}}
Since the magnetic helicity density is a gauge-dependent quantity (see the discussion in Ref.~\cite{Boyer:2025jno}), we should be careful about the local generation of baryons from the primordial magnetic helicity.
Suppose that the magnetic field approximately consists of random blobs of localized magnetic helicity.
Under this assumption, we may generalize the formulae \eqref{eq:nB+L_3} and \eqref{eq:nB+L_1} as if the magnetic helicity is locally gauge-independent by taking the average over a local volume $V$ much larger than the characteristic scale of the blobs.\footnote{
    We assume that the surface contribution is negligible in this procedure.
}
This is valid only when we limit ourselves to discuss baryon density distribution at large scales compared with the magnetic coherence length $\xi_\mathrm{M}$.

With this assumption, we examine the baryon isocurvature constraint on nonhelical ($\vert\epsilon_\mathrm{i}\vert\ll1$) primordial magnetic fields~\cite{Kamada:2020bmb}.
For simplicity,\footnote{For more general and detailed discussion, see Refs.~\cite{Kamada:2020bmb, Boyer:2025jno}.} we assume a monochromatic magnetic power spectrum, $\langle B_{{\rm c}\mkern-7.5mu/i}(\vec k)B_{{\rm c}\mkern-7.5mu/j}(-\vec k)\rangle\propto P_{ij}(\hat k)\delta(k-2\pi/\xi_\mathrm{M,i})$, where $P_{ij}(\hat k)\coloneq\delta_{ij}-\hat k_i\hat k_j$ is the projection operator, and we have omitted the antisymmetric term proportional to $\epsilon_\mathrm{i}$.
This implies that the baryons generated by the helicity decay contribute as the spatial fluctuations of the baryon asymmetry around the homogeneous component $\overline{\eta_B^\mathrm{obs}}$. 
Assuming that baryon isocurvature perturbations are generated as
\begin{equation}
    \delta\eta_{B,\mathrm{fo}}=\dfrac{n_{B,\mathrm{fo}}^\mathrm{diss}+n_{B,\mathrm{fo}}^\mathrm{cdec}}{s_\mathrm{fo}},
\end{equation}
it follows that the baryon power spectrum has an infrared tail at $k\ll 2\pi/\xi_\mathrm{M,i}$ (see Appendix for derivation. We assume Gaussianity for the probability distribution of the magnetic field and compute the spectrum taking the Coulomb gauge).
Namely, 
\begin{align}
    \!\!\!\!\!\left\langle\big(\delta\eta_{B,\mathrm{fo}}(\vec x)\big)^{\!2}\right\rangle
    &\simeq \dfrac{1}{6}S_B^{\mathrm{OE}}\!\left(\dfrac{\xi_\mathrm{M,i}}{2\pi}\right)^{\!\!2}\!\int\dd\ln k \,k^2\!+\cdots\label{eq:IR}\\
    &\simeq \dfrac{1}{3}S_B^{\mathrm{OE}},\label{eq:tot}
\end{align}
where the order estimate $S_B^{\mathrm{OE}}=(\overline{n_{B,\mathrm{fo}}^\mathrm{cdec}}+\overline{n_{B,\mathrm{fo}}^\mathrm{diss}})^2/s^2_\mathrm{fo}$ is given by the halves of Eqs.~\eqref{eq:nB+L_3} and \eqref{eq:nB+L_1} with $\epsilon_\mathrm{i}=1$.

If the characteristic length scale $\xi_\mathrm{M,i}$ is sufficiently large, {\it i.e.}, larger than the neutron diffusion scale at the BBN epoch, $0.0025\,\mathrm{pc}$, the baryon fluctuation \eqref{eq:tot} survives and affect the primordial element abundances, in particular, the deuterium abundance \cite{Kamada:2020bmb}.
On the other hand, even if $\xi_\mathrm{M,i}$ is much smaller than the neutron diffusion scale, the infrared modes $\propto k^2$ \eqref{eq:IR} generated at the electroweak scale survive against the neutron diffusion.\footnote{The $k^2$ behavior is not a universal one, and the power depends on the magnetic power spectrum \cite{Kamada:2020bmb}.}

Therefore, the total baryon isocurvature perturbations that affect the BBN are roughly estimated as
\begin{align}
    \!\!\!\!\!\left\langle\big(\delta\eta_{B,\mathrm{BBN}}(\vec x)\big)^{\!2}\right\rangle
    &\simeq\dfrac{1}{3}S_B^{\mathrm{OE}}\,\mathrm{min}\left\{\!1,\,\dfrac{1}{2}\!\left(\dfrac{\xi_\mathrm{M,i}}{\lambda_\mathrm{neu}}\right)^{\!\!2}\!\right\},
\end{align}
which should be less than $0.016\overline{\eta_B^\mathrm{obs}}^2$ \cite{Inomata:2018htm}. 
This condition excludes the red-shaded regions in {\it Bottom} panel of Fig.~\ref{fig:blazar}.

\noindent\textit{\textbf{Discussion.---\,}}
Finally, let us discuss the baryon asymmetry constraints in light of the blazar observation.
In Figs.~\ref{fig:blazar}, we plot the baryon overproduction/isocurvature exclusions at $T=T_\mathrm{fo}$ by red lines, solid one for $\alpha=1$, dash-dot one for $\alpha=10^{-9}/10^{-10}$, and dotted one for $\alpha=0$.

In {\it Top} panel, the green circles exemplify maximally helical primordial magnetic fields that account for the origin of the BAU: for $\alpha=1$, $B_\mathrm{i}\sim10^{-13}\,\mathrm{G}$ and $\xi_\mathrm{M,i}\sim10^{-18}\,\mathrm{Mpc}$ (solid circle in the left-top) is the possible strongest because the black-shaded region is excluded by the magneto-hydrodynamics (MHD) \cite{Uchida:2024ude};\footnote{This exclusion has not been quantitatively established \cite{Banerjee:2004df,Brandenburg:2017neh,Hosking:2022umv,Uchida:2024ude}, and the lines in the plot show the most conservative ones.} for $\alpha=0$, a causally-generated magnetic field can sit around $B_\mathrm{i}\sim10^{-8}\,\mathrm{G}$ and $\xi_\mathrm{M,i}\sim10^{-10}\,\mathrm{Mpc}$ (dashed one in the left-top); for $\alpha=10^{-9}$, a possibility is $B_\mathrm{i}\sim10^{-16}\,\mathrm{G}$ and $\xi_\mathrm{M,i}\sim10^{-1}\,\mathrm{Mpc}$ (dash-dot one), which is beyond the horizon at $T_\mathrm{fo}$ but can be generated by inflation models.
Since the MHD exclusion is time-dependent and moves rightward until the recombination epoch, the first two examples experience an MHD decay process, in which the magnetic helicity $\propto B^2\xi_\mathrm{M}$ is conserved \cite{frisch75}.
They end up with the right-bottom circles below ($B\sim 10^{-16}\,\mathrm{G}$ and $\xi_\mathrm{M}\sim10^{-10}\,\mathrm{Mpc}$ for the solid one) and above ($B\sim 10^{-10}\,\mathrm{G}$ and $\xi_\mathrm{M}\sim10^{-6}\,\mathrm{Mpc}$ for the dashed one) the blazar lower bound \eqref{eq:blazar}.
The dash-dot circle slightly above the blazar bound is ignorant of the MHD decay process and frozen until the recombination epoch.

In {\it Bottom} panel, the green circles exemplify non-helical primordial magnetic fields that are safe from the baryon isocurvature problem: for $\alpha=1$, $B_\mathrm{i}\sim10^{-10}\,\mathrm{G}$ and $\xi_\mathrm{M,i}\sim10^{-17}\,\mathrm{Mpc}$ (solid circle) is the possible strongest; for $\alpha=0$, a causally-generated magnetic field can sit around $B_\mathrm{i}\sim10^{-8}\,\mathrm{G}$ and $\xi_\mathrm{M,i}\sim10^{-10}\,\mathrm{Mpc}$ (dashed one); for $\alpha=10^{-10}$, a possibility is $B_\mathrm{i}\sim10^{-16}\,\mathrm{G}$ and $\xi_\mathrm{M,i}\sim10^{-1}\,\mathrm{Mpc}$ (dash-dot one), which is beyond the horizon at $T_\mathrm{fo}$.
The first two examples experience an MHD decay process, in which the Hosking integral $\sim B^4\xi_\mathrm{M}^5$ is conserved \cite{Hosking:2020wom}.
They end up with the right-bottom circles below ($B\sim 10^{-17}\,\mathrm{G}$ and $\xi_\mathrm{M}\sim10^{-11}\,\mathrm{Mpc}$ for the solid one) and above ($B\sim 10^{-12}\,\mathrm{G}$ and $\xi_\mathrm{M}\sim10^{-7}\,\mathrm{Mpc}$ for the dashed one) the blazar lower bound \eqref{eq:blazar}.
The dash-dot circle above the blazar bound is ignorant of the MHD decay process and frozen until the recombination epoch.
A caveat is that we treat the gauge-dependent magnetic helicity as if a locally-defined quantity, which can be reasonable only well within the Hubble horizon, where the Cosmic Microwave Background (CMB) and BBN constrain the baryon asymmetry distribution, $\xi_{\rm M}\ll H_0^{-1}$.

To summarize,
it turns out that maximally helical magnetic fields generated before the EWSB generally fails to explain the origin of the void magnetic field if $\alpha\gg10^{-9}$, while there exist magnetic fields that both satisfy the blazar lower bound and explain the origin of the BAU if $\alpha\lesssim10^{-9}$.
As to non-helical magnetic fields generated before the EWSB, there exist an window that is safe against the baryon isocurvature problem and satisfies the blazar lower bound if $\alpha\lesssim10^{-10}$.
We leave investigating the precise value of $\alpha$ as a future work.

\paragraph{Acknowledgements}

We thank Hajime Fukuda and Kohei Kamada for discussion.
This work is supported by MEXT Leading Initiative for Excellent Young Researchers Grant No.~JPMXS0320210099 [YH] and JSPS KAKENHI Grant Nos.~JP24H00976 [YH], JP24K07035 [YH], JP24KF0167 [YH], JP22K14044 [KM], and JP23KJ0642 [FU].
This work is also supported by World Premier International Research Center Initiative (WPI Initiative), MEXT, Japan, and by IBS under the project code IBS-R018-D3.

\small
\bibliographystyle{utphys}
\bibliography{ref}

\appendix

\section{Assumptions on the magnetic field}
\noindent
In this appendix section, we explain the assumptions we have made implicitly in the main text.

First, we model the dissipation of the unconfined magnetic helicity as
\begin{align}
    \label{eq:MHD}
    \dfrac{\dd\big(\kappa^2\vec{A}_{\rm c\mkern-7.5mu/}\cdot\vec{B}_{\rm c\mkern-7.5mu/}\big)}{\dd t}
    &\simeq -2a^{-1}\kappa^2\vec{E}_{\rm c\mkern-7.5mu/}\cdot\vec{B}_{\rm c\mkern-7.5mu/}\notag\\
    &=-2a^{-1}\kappa^2\sigma_{\rm c\mkern-7.5mu/}^{-1}\,\vec{B}_{\rm c\mkern-7.5mu/}\cdot\vec\nabla\times\vec{B}_{\rm c\mkern-7.5mu/},
\end{align}
where we neglect the total derivative in the first similarity, and $\kappa$ takes a different value depending on the assumption.
Namely, we take $\kappa=g'\cos\theta_\mathrm{eff}$ based on the assumption of $\mathrm{U}(1)_Y$ flux conservation, while $\kappa_\mathrm{KL}=1$ based on the assumption of magnetic energy density conservation.
$\sigma_{\rm c\mkern-7.5mu/}$ is the electric conductivity in the comoving unit.
The spatial derivative is taken with respect to the comoving spatial coordinate.

Second, we have parametrized the global parity violation in the following way:
\begin{align}
    \qty[\kappa^2\overline{\vec{A}_{\rm c\mkern-7.5mu/} \cdot \vec{B}_{\rm c\mkern-7.5mu/}}]_\mathrm{fo}
        &=\dfrac{\epsilon_\mathrm{i}}{2\pi}\kappa(T=0)^2 B^2_\mathrm{i}\xi_\mathrm{M\,i},
    \label{eq:AB}\\[5pt]
    \qty[\kappa^2\overline{\vec{B}_{\rm c\mkern-7.5mu/}\cdot\vec\nabla\times\vec{B}_{\rm c\mkern-7.5mu/}}]_\mathrm{fo}
        &=\dfrac{\epsilon_\mathrm{i}}{2\pi}\kappa(T=0)^2 B^2_\mathrm{i}\xi_\mathrm{M,\,i}^{-1},
    \label{eq:BdB}
\end{align}
where $\epsilon_\mathrm{i},\;\vert\epsilon_\mathrm{i}\vert\leq1$ is the so-called magnetic helicity fraction.
For simplicity, we assume the proportionality between the symmetric and asymmetric parts of the magnetic power spectrum, namely $\langle B_{{\rm c\mkern-7.5mu/}i}(\vec k)B_{{\rm c\mkern-7.5mu/}j}(\vec k')\rangle=(2\pi)^3\delta^3(\vec k+\vec k')(P_{ij}(\hat k)P_{\rm c\mkern-7.5mu/}(k)+ \epsilon_{ijk}i\hat k_kP^{\mathrm{asym}}_{\rm c\mkern-7.5mu/}(k))$, where we assume $P^{\mathrm{asym}}_{\rm c\mkern-7.5mu/}(k)=\epsilon P_{\rm c\mkern-7.5mu/}(k)$.
Then the magnetic and the current helicities share the same helicity fraction $\epsilon$.

Third, the monochromatic (and nonhelical) magnetic power spectrum we have introduced is 
\begin{align}
    \left\langle B_{{\rm c}\mkern-7.5mu/i}(\vec k)B_{{\rm c}\mkern-7.5mu/j}(\vec k')\right\rangle
        =&\,(2\pi)^3\delta^3\big(\vec k+\vec k'\big)P_{ij}(\hat k)\notag\\
        &\cdot\dfrac{1}{4}B_\mathrm{i}^2\xi_\mathrm{M,i}^2\,\delta\Big(k-\dfrac{2\pi}{\xi_\mathrm{M,i}}\Big),
    \label{eq:monochromatic}
\end{align}
based on which we can explicitly compute the baryon isocurvature perturbations.

\section{Derivation of Eqs.~\eqref{eq:IR} and \eqref{eq:tot}}
\label{sec:derivation}
\noindent
In this appendix section, we show the derivation of Eqs.~\eqref{eq:tot} and \eqref{eq:IR}.
We have these formulae\footnote{The expression, Eq.~\eqref{eq:formula}, and the values of $C_1$ and $C_2$ are obtained in a similar manner to Eq.~\eqref{eq:diss_net}. Namely, baryons are sourced by the contributions in the first line in Eq.~\eqref{eq:B+L}, out of which the contribution $\propto(\Delta\theta_\mathrm{eff})H_{{\rm c}\mkern-7.5mu/}$ is what we identify as the confined helicity decay (``cdec'') and the contribution $\propto\Delta H_{{\rm c}\mkern-7.5mu/}$ is identified as the helicity dissipation (``diss'') contribution. The ``cdec'' contribution may be canceled by higgs dynamics, so we have introduced an unknown efficiency parameter $\alpha$. Then, locally, the former contribution is $\propto\alpha A_{{\rm c}\mkern-7.5mu/}\cdot B_{{\rm c}\mkern-7.5mu/}$ while the latter is $\propto \vec\nabla\times\vec B_{{\rm c}\mkern-7.5mu/}\cdot\vec B_{{\rm c}\mkern-7.5mu/}$ (See Eqs.~\eqref{eq:nB+L_3} and \eqref{eq:nB+L_1} for the helical case).},
\begin{align}
    \delta\eta_{B,\mathrm{fo}}
    &=\alpha\eta_{B,\mathrm{fo}}^\mathrm{cdec}+\eta_{B,\mathrm{fo}}^\mathrm{diss}\notag\\
    &=2\pi(\alpha C_1 \vec A_{{\rm c}\mkern-7.5mu/}+C_2\vec\nabla\times\vec B_{{\rm c}\mkern-7.5mu/})\cdot\vec B_{{\rm c}\mkern-7.5mu/},
    \label{eq:formula}
\end{align}
where $C_1=5\cdot10^{32}\,\mathrm{G}^2\,\mathrm{Mpc}$ and $C_2=8\cdot10^{13}\,\mathrm{G}^2\,\mathrm{Mpc}^{-1}$, and, working with the Coulomb gauge condition, $\vec\nabla\cdot\vec A_{{\rm c}\mkern-7.5mu/}=0$,
\begin{align}
    \left\langle A_{{\rm c}\mkern-7.5mu/i}(\vec k)B_{{\rm c}\mkern-7.5mu/j}(\vec k')\right\rangle
        &=i\epsilon_{irs}k_rk^{-2}\left\langle B_{{\rm c}\mkern-7.5mu/s}(\vec k)B_{{\rm c}\mkern-7.5mu/j}(\vec k')\right\rangle,\\
    \left\langle A_{{\rm c}\mkern-7.5mu/i}(\vec k)A_{{\rm c}\mkern-7.5mu/j}(\vec k')\right\rangle
        &=k^{-2}\left\langle B_{{\rm c}\mkern-7.5mu/i}(\vec k)B_{{\rm c}\mkern-7.5mu/j}(\vec k')\right\rangle,\\
    \left\langle j_{{\rm c}\mkern-7.5mu/i}(\vec k)A_{{\rm c}\mkern-7.5mu/j}(\vec k')\right\rangle
        &=\left\langle B_{{\rm c}\mkern-7.5mu/i}(\vec k)B_{{\rm c}\mkern-7.5mu/j}(\vec k')\right\rangle,\\
    \left\langle j_{{\rm c}\mkern-7.5mu/i}(\vec k)B_{{\rm c}\mkern-7.5mu/j}(\vec k')\right\rangle
        &=i\epsilon_{irs}k_r\left\langle B_{{\rm c}\mkern-7.5mu/s}(\vec k)B_{{\rm c}\mkern-7.5mu/j}(\vec k')\right\rangle,\\
    \left\langle j_{{\rm c}\mkern-7.5mu/i}(\vec k)j_{{\rm c}\mkern-7.5mu/j}(\vec k')\right\rangle
        &=k^2\left\langle B_{{\rm c}\mkern-7.5mu/i}(\vec k)B_{{\rm c}\mkern-7.5mu/j}(\vec k')\right\rangle,
\end{align}
where $\vec j_{{\rm c}\mkern-7.5mu/}\coloneq\vec\nabla\times\vec B_{{\rm c}\mkern-7.5mu/}$ and we hereafter substitute Eq.~\eqref{eq:monochromatic}.

The power spectrum of the baryon isocurvature perturbations becomes
\begin{align}
    &\left\langle\delta\eta_{B,\mathrm{fo}}(\vec k)\delta\eta_{B,\mathrm{fo}}(\vec k') \right\rangle\notag\\
        &\qquad=4\pi^2(\alpha^2C_1^2T_1+\alpha C_1C_2(T_2+T_3)+C_2^2T_4),
\end{align}
where each contribution is computed by assuming Gaussianity (first similarity in each equation below), neglecting terms proportional to $\epsilon_\mathrm{i}$ (second similarity), and using these notations, $\Theta(x\leq0)=0$, $\Theta(x>0)=1$, and $\int_{\vec{p},\vec{q}}\coloneq\int\!\frac{\dd^3p}{(2\pi)^3}\int\!\frac{\dd^3q}{(2\pi)^3}$,
\begin{align}
    T_1
        \coloneq&\left\langle\vec A_{{\rm c}\mkern-7.5mu/}\cdot \vec B_{{\rm c}\mkern-7.5mu/}(\vec k)\,\vec A_{{\rm c}\mkern-7.5mu/}\cdot \vec B_{{\rm c}\mkern-7.5mu/}(\vec k')\right\rangle\notag\\
        \simeq&\left\langle\vec A_{{\rm c}\mkern-7.5mu/}\cdot \vec B_{{\rm c}\mkern-7.5mu/}(\vec k)\right\rangle\!\left\langle\vec A_{{\rm c}\mkern-7.5mu/}\cdot \vec B_{{\rm c}\mkern-7.5mu/}(\vec k')\right\rangle\notag\\
        &\,+\int_{\vec p,\vec q}\!\!\left\langle A_{{\rm c}\mkern-7.5mu/i}(\vec p)A_{{\rm c}\mkern-7.5mu/j}(\vec q)\right\rangle\!\left\langle B_{{\rm c}\mkern-7.5mu/i}(\vec k-\vec p) B_{{\rm c}\mkern-7.5mu/j}(\vec k'\!-\vec q)\right\rangle\notag\\
        &\,+\int_{\vec p,\vec q}\!\!\left\langle A_{{\rm c}\mkern-7.5mu/i}(\vec p)B_{{\rm c}\mkern-7.5mu/j}(\vec k'\!-\vec q)\right\rangle\!\left\langle A_{{\rm c}\mkern-7.5mu/j}(\vec q) B_{{\rm c}\mkern-7.5mu/i}(\vec k-\vec p)\right\rangle\notag\\
        \simeq&\,(2\pi)^3\delta^3\big(\vec k+\vec k'\big)\qty(\dfrac{1}{4}B_{\mathrm i}^2\xi_\mathrm{M,i}^2)^{\!\!2}\notag\\
        &\cdot\dfrac{1}{\pi^2k}\Theta\qty(2\cdot\!\dfrac{2\pi}{\xi_\mathrm{M,i}}-k)\bigg(\qty(\dfrac{k\xi_\mathrm{M,i}}{4\pi})^{\!\!2}-1\bigg)^{\!\!2},\\[6pt]
    T_2
        \coloneq&\left\langle\vec A_{{\rm c}\mkern-7.5mu/}\cdot \vec B_{{\rm c}\mkern-7.5mu/}(\vec k)\,\vec j_{{\rm c}\mkern-7.5mu/}\cdot \vec B_{{\rm c}\mkern-7.5mu/}(\vec k')\right\rangle\notag\\
        \simeq&\left\langle\vec A_{{\rm c}\mkern-7.5mu/}\cdot \vec B_{{\rm c}\mkern-7.5mu/}(\vec k)\right\rangle\!\left\langle\vec j_{{\rm c}\mkern-7.5mu/}\cdot \vec B_{{\rm c}\mkern-7.5mu/}(\vec k')\right\rangle\notag\\
        &\,+\int_{\vec p,\vec q}\!\!\left\langle A_{{\rm c}\mkern-7.5mu/i}(\vec p)j_{{\rm c}\mkern-7.5mu/j}(\vec q)\right\rangle\!\left\langle B_{{\rm c}\mkern-7.5mu/i}(\vec k-\vec p) B_{{\rm c}\mkern-7.5mu/j}(\vec k'\!-\vec q)\right\rangle\notag\\
        &\,+\int_{\vec p,\vec q}\!\!\left\langle A_{{\rm c}\mkern-7.5mu/i}(\vec p)B_{{\rm c}\mkern-7.5mu/j}(\vec k'\!-\vec q)\right\rangle\!\left\langle j_{{\rm c}\mkern-7.5mu/j}(\vec q) B_{{\rm c}\mkern-7.5mu/i}(\vec k-\vec p)\right\rangle\notag\\
        \simeq&\left(\dfrac{2\pi}{\xi_\mathrm{M,i}}\right)^{\!\!2}T_1,\\[6pt]
    T_3
        \coloneq&\left\langle\vec j_{{\rm c}\mkern-7.5mu/}\cdot \vec B_{{\rm c}\mkern-7.5mu/}(\vec k)\,\vec A_{{\rm c}\mkern-7.5mu/}\cdot \vec B_{{\rm c}\mkern-7.5mu/}(\vec k')\right\rangle\notag\\
        =&T_2,\\[6pt]
    T_4
        \coloneq&\left\langle\vec j_{{\rm c}\mkern-7.5mu/}\cdot \vec B_{{\rm c}\mkern-7.5mu/}(\vec k)\,\vec j_{{\rm c}\mkern-7.5mu/}\cdot \vec B_{{\rm c}\mkern-7.5mu/}(\vec k')\right\rangle\notag\\
        \simeq&\left\langle\vec j_{{\rm c}\mkern-7.5mu/}\cdot \vec B_{{\rm c}\mkern-7.5mu/}(\vec k)\right\rangle\!\left\langle\vec j_{{\rm c}\mkern-7.5mu/}\cdot \vec B_{{\rm c}\mkern-7.5mu/}(\vec k')\right\rangle\notag\\
        &\,+\int_{\vec p,\vec q}\!\!\left\langle j_{{\rm c}\mkern-7.5mu/i}(\vec p)j_{{\rm c}\mkern-7.5mu/j}(\vec q)\right\rangle\!\left\langle B_{{\rm c}\mkern-7.5mu/i}(\vec k-\vec p) B_{{\rm c}\mkern-7.5mu/j}(\vec k'\!-\vec q)\right\rangle\notag\\
        &\,+\int_{\vec p,\vec q}\!\!\left\langle j_{{\rm c}\mkern-7.5mu/i}(\vec p)B_{{\rm c}\mkern-7.5mu/j}(\vec k'\!-\vec q)\right\rangle\!\left\langle j_{{\rm c}\mkern-7.5mu/j}(\vec q) B_{{\rm c}\mkern-7.5mu/i}(\vec k-\vec p)\right\rangle\notag\\
        \simeq&\left(\dfrac{2\pi}{\xi_\mathrm{M,i}}\right)^{\!\!4}T_1.
\end{align}

In particular, the infrared mode of the baryon isocurvature perturbations therefore scales as
\begin{align}
    &\left\langle\delta\eta_{B,\mathrm{fo}}(\vec k)\delta\eta_{B,\mathrm{fo}}(\vec k') \right\rangle\notag\\
        &\qquad\simeq\bigg(\!\alpha C_1+\left(\dfrac{2\pi}{\xi_\mathrm{M,i}}\right)^{\!\!2}C_2\!\bigg)^{\!\!2}4\pi^2T_1\notag\\
        &\qquad\simeq\qty(\dfrac{1}{4}B_{\mathrm i}^2\xi_\mathrm{M,i}^2)^{\!\!2}\bigg(\!\alpha C_1+\left(\dfrac{2\pi}{\xi_\mathrm{M,i}}\right)^{\!\!2}C_2\!\bigg)^{\!\!2}\notag\\
        &\qquad\quad\cdot(2\pi)^3\delta^3\big(\vec k+\vec k'\big)\dfrac{4}{k}(1+\cdots),
\end{align}
and hence
\begin{align}
    &\left\langle\big(\delta\eta_{B,\mathrm{fo}}(\vec x)\big)^{\!2}\right\rangle\notag\\
        &\qquad\simeq\qty(\dfrac{1}{4}B_{\mathrm i}^2\xi_\mathrm{M,i}^2)^{\!\!2}\bigg(\!\alpha C_1+\left(\dfrac{2\pi}{\xi_\mathrm{M,i}}\right)^{\!\!2}C_2\!\bigg)^{\!\!2}\notag\\
        &\qquad\quad\cdot\dfrac{2}{\pi^2}\int\dd k \,k(1+\cdots)\label{eq:IRtail}\\
        &\qquad\simeq\dfrac{1}{3}\qty(B_{\mathrm i}^2\xi_\mathrm{M,i})^2\bigg(\!\alpha C_1+\left(\dfrac{2\pi}{\xi_\mathrm{M,i}}\right)^{\!\!2}C_2\!\bigg)^{\!\!2}.\label{eq:total}
\end{align}

\section{Input values in Eq.~\eqref{eq:diss_net}}
\label{sec:suppl}
\noindent
In this appendix section, we explain the necessary information to obtain Eq.~\eqref{eq:diss_net}.

We used the $Z$-pole values of Standard Model parameters \cite{ParticleDataGroup:2024cfk},
\begin{align}
    g=0.6,\qquad
    g'=0.3,\qquad
    \sin\theta_\mathrm{w}=0.5,
\end{align}
the standard expansion history of the universe \cite{Mukhanov:2005sc} at the spharelon freezeout temperature,
\begin{align}
    a_\mathrm{fo}&=(3.91/106.75)^{1/3}(2.73\,\mathrm{K}/T_\mathrm{fo}),\\
    H_\mathrm{fo}&=-T_\mathrm{fo}^{-1}(\dd T/\dd t)_\mathrm{fo}=\sqrt{106.75/90}\,\pi T_\mathrm{fo}^2/M_\mathrm{Pl},\\
    s_\mathrm{fo}&=(2\pi^2/45)106.75T_\mathrm{fo}^3,
\end{align}
and the approximate conductivity $\sigma_{\rm c\mkern-7.5mu/}(T_\mathrm{fo})\sim10^2a_\mathrm{fo}T_\mathrm{fo}$ \cite{Kamada:2016cnb}, where $M_\mathrm{Pl}=2.44\times 10^{18}\,\mathrm{GeV}$ is the reduced Planck mass.
As we have worked in the Heaviside--Lorentz unit, we have $1\,\mathrm{G}=1.95\times10^{-20}\,\mathrm{GeV}^2$ and $1\,\mathrm{Mpc}=1.56\times10^{38}\,\mathrm{GeV}^{-1}$.

In addition, $\theta_\mathrm{eff}$ and $\dot \theta_\mathrm{eff}$ at the sphaleron freezeout temperature are determined by the formula \eqref{eq:coseff} with the empirical formula \cite{Kamada:2016cnb}
\begin{align}
    v_3(T)
        \simeq0.23\sqrt{T\qty(162-\dfrac{T}{1\,\mathrm{GeV}})},
\end{align}
which accurately reproduces the lattice result at $T>140\,\mathrm{GeV}$ \cite{DOnofrio:2015gop} and is quite a good approximation of the perturbative computation down to $T\sim130\,\mathrm{GeV}$ \cite{Hamada:2025cwu}.

\section{Evolution of magnetic fields}
\noindent
We explain how we take into account the evolution of magnetic field strength and coherence length, trying to avoid confusion in distinguishing the comoving quantities and the present-day quantities.

In our discussion in the main text, we introduce the comoving magnetic field strength $B$ and the comoving magnetic coherence length $\xi_\mathrm{M}$, which are obtained by rescaling the physical quantities by powers of the scale factor (See, \textit{e.g.}, Ref.~\cite{RoperPol:2025lgc} for a pedagogical review).
Namely,
\begin{align}
    B
    =a^2 B_\mathrm{physical},\quad
    \xi_\mathrm{M}
    =a^{-1}\xi_\mathrm{M,\,physical}.
\end{align}

The pure electromagnetism is conformal invariant, and hence the expansion of the universe does not affect $B$ nor $\xi_{\mathrm M}$.
In this sense, the comoving quantities are often interpreted as their present-day values, by taking the present-day scale factor to be unity, $a_0=1$.

However, once we take into account charged particles that couples to the electromagnetism, apart from the cosmic expansion, the magnetic field is advected, stretched, compressed, and dissipated by the plasma fluid.
These effects are described by the MHD equations, which governs the non-trivial time dependence of the comoving quantities such as $B$ and $\xi_\mathrm{M}$.
In this sense, generally speaking, comoving values are not necessarily equal to the present-day values.

The MHD evolution of the cosmological magnetic field has been extensively studied in the literature \cite{Banerjee:2004df,Brandenburg:2017neh,Hosking:2022umv,Uchida:2024ude}.
To find relations between the initial comoving parameters $B_\mathrm{i}$ and $\xi_\mathrm{M,\,i}$ and their present-day values $B$ and $\xi_\mathrm{M}$, we have considered two conditions \cite{Banerjee:2004df, Uchida:2024ude}.

One is the conservation laws (the green arrows in Fig.~\ref{fig:blazar}), namely
\begin{align}
    \begin{cases}
        B_\mathrm{i}^2\xi_\mathrm{M,\,i}=B^2\xi_\mathrm{M}&\text{if maximally helical }(\epsilon_{\mathrm i}=\pm1)\text{ \cite{frisch75}}\\
        B_\mathrm{i}^4\xi_\mathrm{M,\,i}^5=B^4\xi_\mathrm{M}^5&\text{if non-helical }(\vert\epsilon_{\mathrm i}\vert\ll1)\text{ \cite{Hosking:2020wom}}.
    \end{cases}
\end{align}
The former one has been recognized in the standard literature \cite{Durrer:2013pga}, and the latter one was proposed quite recently \cite{Hosking:2020wom} and is more or less established in dedicated numerical studies \cite{Zhou:2022xhk,Brandenburg:2025ccv}.

The other condition is the nonlinear decay timescale, which should be shorter than the cosmological time for the magnetic field to be processed by MHD dynamics.
Generally, magnetic fields with shorter coherence length have smaller decay time scale, implying that, at a given time in the history of the universe, MHD has already processed magnetic fields of shorter coherence length (black-hatched exclusion in Fig.~\ref{fig:blazar}).
However, note that the analytic estimate of decay timescale has not been established yet \cite{Banerjee:2004df,Brandenburg:2017neh,Hosking:2022umv,Uchida:2024ude}.

\section{Zoom-up figures}
\begin{figure}[ht]
    \includegraphics[keepaspectratio, width=.4\textwidth]{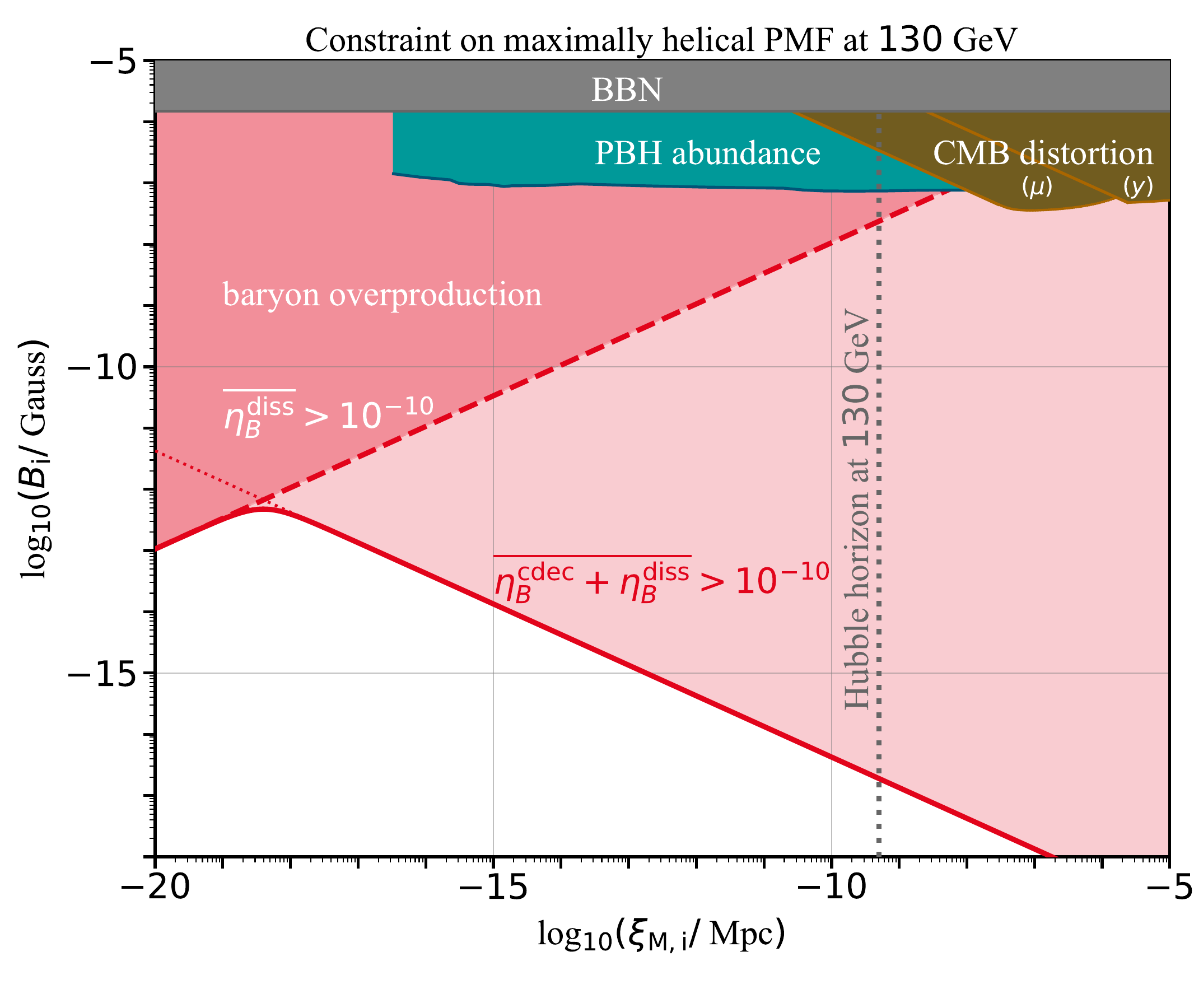}
    \caption{The baryon-overproduction constraint for maximally helical ($\epsilon_\mathrm{i}=1$) primordial magnetic field at $T_\mathrm{fo}\simeq130\,\mathrm{GeV}$. The comoving coherence length $\xi_\mathrm{M,i}$ and the comoving magnetic field strength $B_\mathrm{i}$ in the red shaded region above the dashed line violates Eq.~\eqref{eq:BO} with $\alpha=0$ and those in the red region above the solid line violates Eq.~\eqref{eq:BO} with $\alpha=1$. For reference, the gray dotted line is the comoving Hubble horizon, the gray-shaded exclusion is the BBN bound \cite{Kawasaki:2012va}, the teal-shaded exclusion is the primordial black hole (PBH) abundance bound \cite{Kushwaha:2024zhd}, and the brown-shaded is the CMB distortion bound \cite{Uchida:2024iog}.}
    \label{fig:BO}
\end{figure}

\begin{figure}[ht]
    \includegraphics[keepaspectratio, width=.4\textwidth]{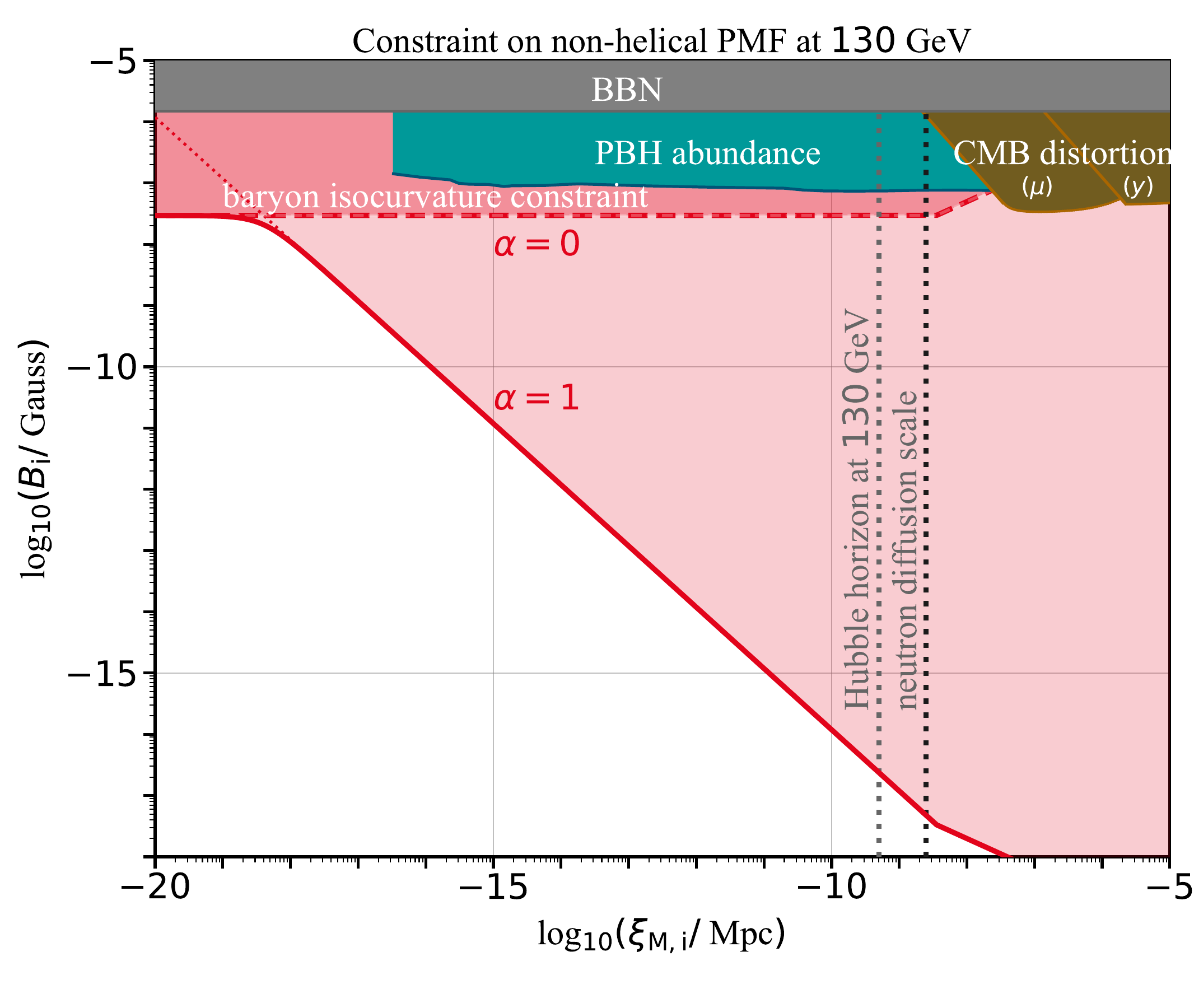}
    \caption{The baryon isocurvature constraint for non-helical ($\vert\epsilon_\mathrm{i}\vert\ll1$) primordial magnetic field at $T_\mathrm{fo}\simeq130\,\mathrm{GeV}$. The comoving coherence length $\xi_\mathrm{M,i}$ and the comoving magnetic field strength $B_\mathrm{i}$ in the red shaded regions above the dashed (solid) line result in too much baryon isocurvature perturbations with $\alpha=0$ ($\alpha=1$, {\it resp.}).  
    For reference, the dense-gray dotted line is the neutron diffusion scale $\lambda_\mathrm{neu}$.
    See also Fig.~\ref{fig:BO}.
    }
    \label{fig:BI}
\end{figure}

\noindent
Here we show zoom-up figures (same as the exclusion in Fig.~\ref{fig:blazar}) that excludes primordial magnetic fields at $130$ GeV by the baryon asymmetry problems, namely the baryon overproduction for Fig.~\ref{fig:BO} and the baryon isocurvature problem for Fig.~\ref{fig:BI}.


\end{document}